# Recent Advances in Graphene-Based Pressure Sensors: A Review

Zhe Zhang, Quan Liu, Hongliang Ma, Ningfeng Ke, Jie Ding, Wendong Zhang, Xuge Fan

*Abstract*—In recent years, pressure sensors have been widely used as crucial technology components in industrial, healthcare, consumer electronics, and automotive safety applications. With the development of intelligent technologies, there is a growing demand for pressure sensors with higher sensitivity, smaller size, and wider detection range. Graphene and its derivatives, as novel emerging materials in recent years, have received widespread attention from researchers due to their unique mechanical and electrical properties, and are considered as promising sensing materials for the high-performance pressure sensors. In general, graphene-based pressure sensors can be classified into flexible pressure sensors and gas pressure sensors. In this paper, we firstly introduce the basic properties of graphene and its derivatives and then review the research progress of both graphene-based flexible pressure sensors and graphene-based gas pressure sensors respectively, focusing on different sensing mechanisms. Finally, the application prospects of graphene-based pressure sensors as well as future challenges are discussed.

*Index Terms*—graphene; gas pressure sensors; flexible pressure sensors; suspended graphene; graphene and its derivatives

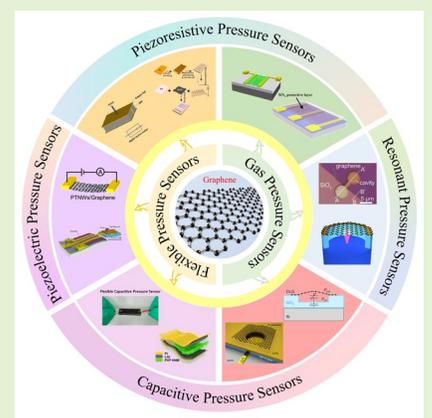

## I. INTRODUCTION

WITH the development of science and technology, pressure sensors have been more and more commonly used in various fields such as consumer electronics[1], [2], medical and health monitoring[3], [4], [5], [6], [7], wearable technology[3], [8], [9], [10], automotive electronics[1], [11], industrial production[12] and aerospace[13], [14] etc. At the same time, with the rapid advancement of the Internet of Things (IoT) and intellectual technologies in recent years, there is an increasing demand for pressure sensors with higher sensitivity, wider detection range, shorter response time and smaller size. It is more and more difficult for the traditional pressure sensors to meet the above demand in various emerging application scenarios. Therefore, it is necessary to utilize the emerging materials such as graphene and its derivatives as the pressure sensing membranes for next generation pressure sensors.

Pressure sensors can be generally classified into gas pressure sensors, liquid pressure sensors and contact pressure sensors according to the different measurement mediums, and according to the current research reports, there are few graphene-based liquid pressure sensors. Most of the researches generally focus on the two kinds of pressure, gas pressure and contact pressure. Gas pressure refers to the average collision force of gas molecules per unit area on the surface of action, which is widely used in industrial control, aerospace and meteorology. Further, there is a special type of gas pressure named acoustic pressure[15], [16], whose pressure change is caused by the propagation of acoustic waves in the gas, and it has important applications in several areas such as microphones, loudspeakers, etc. For contact pressure, with the rapid development of wearable electronic devices, flexible pressure sensors have attracted much attention due to their unique flexibility and comforts. Therefore, in this paper, we mainly review the research progress of pressure sensors in two categories: gas pressure and flexible pressure. For both types of pressure sensors, the core functional component is the sensing membrane, which is responsible for converting the applied pressure signal into an electrical signal and mainly determines the performance of the pressure sensors.

For gas pressure sensors, sensing membranes of most traditional gas pressure sensors are based on silicon materials[17]. Although the manufacturing process of pressure sensors based on silicon sensing membranes is more mature and less costly, silicon-based pressure sensors also face some technical bottlenecks. For instance, silicon-based materials have limitations in terms of mechanical strength, sensitivity to temperature changes, and membrane's thickness. Silicon-based pressure sensors perform poorly in terms of stability and reliability in extreme environments. Furthermore, the rigidity and brittleness of conventional silicon-based pressure sensors

This work was supported by the National Natural Science Foundation of China (62171037 and 62088101), Beijing Natural Science Foundation (4232076), 173 Technical Field Fund (2023-JCJQ-JJ-0971), National Key Research and Development Program of China (2022YFB3204600), Beijing Institute of Technology Science and Technology Innovation Plan, National Science Fund for Excellent Young Scholars (Overseas), Beijing Institute of Technology Teli Young Fellow Program (2021TLQT012).

Please see the Acknowledgment section of this article for the author affiliations.

limit their application in wearable devices. Therefore, researchers are exploring emerging sensing materials such as 2D materials to develop high-performance pressure sensors to meet the growing demand for pressure measurements in modern technology.

Flexible pressure sensors have excellent tensile properties, high sensitivity, and fast response, and thereby are widely used in wearable devices[18], [19], electronic skin[20], [21], human health monitoring[7], [22], [23], [24], [25], motion detection[5], [26], [27], [28], [29], [30], intelligent robotics[31], [32], [33], [34], [35], [36], [37], energy storage devices[38], [39], [40], and biomedical diagnostics[4], [41], [42], [43], [44]. Moreover, with the rapid progress of intelligent technology, there are increasingly high requirements for the performance of flexible pressure sensors. At present, there are two main methods to improve the performance of flexible pressure sensors. One is to build microstructures in dielectrics and electrodes, another is to utilize the novel sensing materials of the flexible pressure sensors, which enhances the performance of the pressure sensors by utilizing the inherent properties of the new materials.

Microstructures are normally classified as single and multiple microstructures. Single microstructures indicate that the sensor contains only one shape of structure, including microstructure patterns such as micro-pyramids[45], [46], [47], micro-domes[48], [49], and micro-pillars[50], [51]. Although microstructure patterns can detect small external mechanical stimuli, when under high pressure, the microstructures can become saturated due to their limited deformation capacity, leading to the failure in sensing high pressure. At the same time, fabricating these microstructure patterns requires complex processes, such as conventional photolithography, electron beam evaporation, and etching, which may significantly increase the manufacturing cost and time, and make it difficult to manufacture in large scale. Multiple microstructures refer to the combination of different shapes of microstructures or layered identical microstructures, which combine the advantages of different microstructures to make flexible pressure sensors with high sensitivity and wide sensing range. However, their preparation strategies are complex and require a high level of equipment[52]. Furthermore, there is another type of microstructures named porous microstructures, such as aerogels, in which the presence of micropores provides high compressibility between the conducting materials as well as large contact variations, and thus the sensors have an ultra-wide pressure detection range. However, the main challenge at present is the lack of control over the pore size and structure, which therefore affects the repeatability and tunability of the sensors[53].

In addition to the use of microstructures to enhance the performance of flexible pressure sensors, another way to improve the performance is to utilize novel sensing materials, and so far, various sensing materials, such as silver nanowires[27], conductive hydrogels[54], graphene[55], MXene[56], [57], [58], [54], [55], and carbon nanotubes[59], [60], [61], [62], [63], have been widely used to fabricate novel flexible pressure sensors. Among the above materials, graphene is not only the most promising candidate for flexible pressure sensors but also has important contributions in the biomedical field[64], [65], [66], [67]. Although all other materials have their own advantages, there are still many problems and shortcomings in some aspects. Firstly, the ultra-high conductivity of graphene makes it more suitable as a sensing material than silver nanowires and MXene. Secondly, in terms of environmental stability, conductive hydrogels, silver nanowires and MXene are highly susceptible to environmental influences, leading to performance degradation. But graphene has good chemical stability and is not easily oxidized or degraded, which will greatly improve the durability and stability of the sensor. Moreover, the preparation process of high-quality MXene is complex, and the manufacturing process is difficult to control and high cost. In contrast, the preparation technology of graphene is relatively more mature, and it has the potential for large-scale production. In addition, the high mechanical strength and Young's modulus of graphene ensure the best flexibility and stretchability, which is more suitable for the fabrication of wearable devices.

This review summarizes the latest research progress of pressure sensors based on graphene and its derivatives (Fig. 1). Firstly, the basic properties of graphene and its derivatives are provided. Second, the latest research progress of graphene-based flexible pressure sensors and gas pressure sensors is summarized according to sensing mechanisms, and their performances are compared. Finally, the existing challenges, technological difficulties, and future prospects of the graphene-based pressure sensors are highlighted.

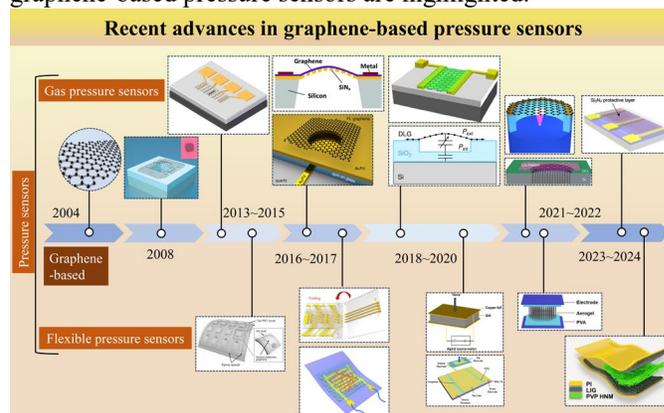

Fig. 1.  Schematic diagram of the progress of graphene-based pressure sensors.

## II. Basic Properties of Graphene and its Derivatives

Among the emerging 2D materials, graphene is one of the most promising 2D materials for pressure sensors due to its unique mechanical and electrical properties, which allows it to be used as an ultra-sensitive membrane to detect various physical quantities including pressure. In recent years, apart from using graphene as sensing materials for sensors, numerous researchers have used derivatives of graphene and its composites with other materials as sensing materials for high-performance sensors[68]. There are various types of graphene derivatives such as graphene oxide (GO)[69], [70], reduced graphene oxide (rGO)[71], [72]. In this section, the basic properties of graphene and its typical derivatives are introduced, which would be useful to better understand the critical role these materials play in the development of pressure sensors. As shown in TABLE I, we present a more comprehensive perspective to summarize the properties of graphene and its derivatives and their implications for pressure sensor.

TABLE I
PROPERTIES OF GRAPHENE AND ITS DERIVATIVES AND THEIR IMPLICATIONS FOR PRESSURE SENSORS.

| | Conductivity and its impact on pressure sensors | | Mechanical properties and their impact on pressure sensors | | The impact of other characteristics on pressure sensors |
|---|---|---|---|---|---|
| Graphene | $10^6$ S·m$^{-1}$ | Extremely sensitive to pressure. | Young's modulus (1 TPa) | Wide measurement range. | Superior adhesion to SiO$_2$ substrates. Ensures strong bonding between graphene and substrate. Impermeability to gas molecules. Long lifetime. |
| GO | Virtually non-conductive | Ideal as a dielectric layer for capacitive pressure sensors. | High compressive strength | Durability and stability. | Ultra-high specific surface area. Provides more active sites that can capture smaller pressure changes. High porosity. Large pressure range. |
| rGO | 200-35100 S m$^{-1}$ | Sensitive to pressure changes. | Young's modulus (200~600 GPa) | High sensitivity and wide measurement range. | Adjustable bandgap. Optimizes the electrical response of the sensor. |

### A. Graphene

Graphene was first successfully isolated from a block of graphite in 2004 by Andre Geim and Konstantin Novoselov, using a simple mechanical exfoliation technique[73]. The stripped atomical layers of graphene are shown in Fig. 2a. It is a hexagonal honeycomb lattice structure formed by a single layer of carbon atoms bonded by sp$^2$ hybridization. This chemical bonding determines that graphene itself has excellent mechanical and electrical properties, including atomic-level thickness (~0.33nm)[74], ultra-high Young's modulus (~ 1TPa)[75], high fracture strength (~125 GPa)[75], high electron mobility (up to $2 \times 10^5$ cm$^2$ V$^{-1}$ s$^{-1}$)[76], and a large surface area (2630 m$^2$/g)[77]. And it is these attractive properties that make it possible to prepare high-performance sensors[78], [79], [80], [81], [82], [83], [84], [85], [86], [87], [88], [89].

Firstly, the most critical reason for utilizing graphene as a sensing material for pressure sensors is its excellent mechanical properties. In 2009, Wang et al. investigated the wrinkling behavior of graphene under external stresses using nanoindentation experiments and molecular dynamics simulations[90], which measured graphene's mechanical properties including bending stiffness and in-plane stiffness. Bending stiffness represents the ability of graphene to resist bending perpendicular to its plane while in-plane stiffness represents the ability of graphene to resist tensile (or compressive) stresses when subjected to tensile (or compressive) stresses in a direction parallel to its surface. It has shown that graphene exhibits extremely high bending stiffness and in-plane stiffness, both of which are important for its application in pressure sensors.

Besides that, the related properties between graphene and gases indicate that graphene is suitable as a sensing material for gas pressure sensors. In 2007, the laboratory of Schedin F showed that graphene membranes can adsorb gas molecules, and the adsorbed gas molecules can change the carrier concentration in graphene, which resulted in a step change in graphene's electrical resistance[91]. As shown in Fig. 2b, this provides an essential reference for the subsequent study of the piezoresistive properties of graphene and the use of graphene as a sensing membrane for gas pressure sensors. From the underlying principle, the external force that is applied to the graphene membrane will lead to a distortion of the hexagonal lattice structure of the carbon atoms arranged in graphene. This distortion will change the bond lengths and angles of the carbon atoms[92], [93], which will change the energy-band structure and lead to a change in the electronic properties of graphene. The pressure can then be measured based on the resistance change of graphene.

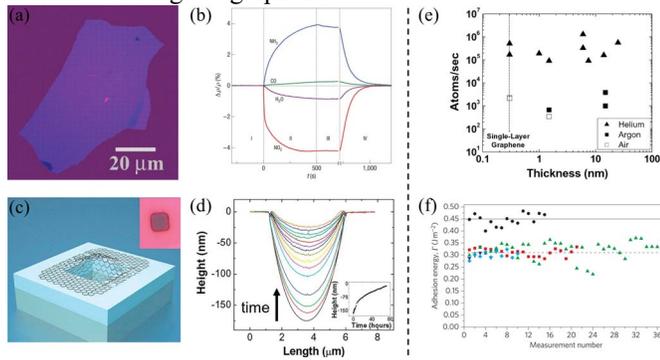

Fig. 2. Fundamental properties of graphene. (a) Photograph of a multilayer graphene sheet on a silicon oxide substrate. (b) Effects of different gas environments on the resistivity of graphene. (c) Schematic diagram of graphene-sealed cavity. (d) AFM line traces taken through the center of the graphene membrane. (e) Scatter plot between gas leakage rate and graphene membrane thickness. (f) Plot of adhesion energy between graphene membrane and $SiO_2$. Adapted from [73], [91], [94] and [95] under a Creative Commons license.

Furthermore, the single-layer graphene membrane is impermeable to many gases (including helium)[94], [96], and the graphene-sealed cavity is illustrated in Fig. 2c. The deflection of the center of the graphene membrane decreases with time, which indicates that there is a slow gas leakage in the cavity, as demonstrated in Fig. 2d. At the same time, the researchers analyzed the relationship between the rate of gas leakage and the thickness of the graphene membrane. According to Fig. 2e, it can be concluded that the rate of gas leakage is independent of the thickness of the graphene, which also demonstrated that the gas leakage in the cavity did not occur through the graphene membrane. Hence, it is concluded that graphene membrane is impermeable to gases and can be used as a sensing membrane material for highly sensitive pressure sensors. In 2011, Steven P. Koenig et al. studied the adhesion energy between graphene membranes and $SiO_2$ substrates[95], in which the adhesion energy between one to five atomical layers of graphene membranes and $SiO_2$ was measured, as shown in Fig. 2f. The experimental results showed that the adhesion energy between a single layer of graphene and $SiO_2$ surface was $0.45 \pm 0.02$ J/m$^2$, while the adhesion energy between two to five layers of graphene and $SiO_2$ surface was $0.31 \pm 0.03$ J/m$^2$. These values are much larger than the adhesion energy in common micromechanical structures, thereby demonstrating the strong adhesion between graphene and $SiO_2$ substrates and graphene is well-suited for applications of pressure sensors.

## B. Graphene Oxide(GO)

GO is formed by introducing oxygen-containing functional groups (e.g., hydroxyl, carboxyl, and epoxy groups) into the graphene layer through an oxidation process, which provides GO some unique properties. Schematic of the oxygen-containing functional groups in graphene oxide is shown in Fig. 3a[97].

Generally, in the fabrication process of graphene-based flexible pressure sensors, a sponge network with a porous structure needs to be formed to achieve a sensitive response to pressure changes, and the three-dimensional porous structure made of graphene oxide exhibits extremely excellent properties[98]. The SEM image of lateral cross-section of GO sponge is shown in Fig. 3b, with an ultra-high specific surface area (2630 m$^2$/g), high porosity, and high compressive strength (~320 kPa). Specifically, the three-dimensional porous structure of graphene oxide foam provides it with an extremely high specific surface area and high porosity, which not only enhances its sensitivity to pressure changes but also provides more active sites and thus facilitates the enhancement of the sensitivity of the pressure sensor.

Furthermore, graphene oxide has several other properties that contribute to its essential role in flexible capacitive pressure sensors. The presence of oxygen-containing functional groups and a large number of defects hinders the movement of electrons in the plane of GO, resulting in GO appearing almost insulating (conductivity of 0.5 S/m depending on the degree of oxidation) and resulting in GO exhibiting a high relative permittivity (in the order of $10^4$ over the frequency range of $0.1 \sim 70$ Hz), as illustrated in Fig. 3c[99]. This is suitable for the fabrication of three-dimensional porous GO foams, making GO an ideal dielectric layer for preparing capacitive pressure sensors[100], [101], [102], [103].

In addition, these oxygen-containing functional groups of GO make it quite chemically active. Consequently, it is easy to efficiently reduce and functionalize it to form reduced graphene oxide (rGO) with good electrical conductivity. Although many of GO's properties are inferior to graphene's properties, it is relatively easy to prepare and is widely used in fields such as flexible pressure sensors and other electronic products.

## C. Reduced Graphene Oxide(rGO)

Reduced graphene oxide is formed by removing or reducing the oxygen functional groups in graphene oxide by some reduction methods. Common reduction methods include chemical reduction[104], [105], [106], thermal reduction[107], [108] and photoreduction[109]. Different reduction methods have an important impact on the properties of reduced graphene oxide. For example, in 2015, Velra et al. explored the impact of the reduction process of graphene oxide on its electrical conductivity, and found that graphene oxide reduced with hydriodic acid exhibited the highest electrical conductivity (103.3 S cm$^{-1}$)[110] and had better flexibility compared to rGO that was reduced using hydrobromic acid and hydrazine compounds (Fig. 3d). This creates favorable conditions for the preparation of flexible pressure sensors. However, compared to chemical reduction that may introduce impurities, thermal reduction avoids this drawback and the degree of reduction of graphene oxide can be controlled by simply adjusting the temperature[111]. In 2019, Wang et al. developed a method to prepare reduced graphene oxide at very high temperatures by Joule heating technique and explored the impact of reduction temperature on the electronic properties of reduced graphene oxide[112]. It was shown that oxygen-containing functional groups were gradually removed with increasing reduction temperature, which resulted in a gradual decrease in the carrier density and a gradual increase in the electrical conductivity. The longitudinal resistivity of the rGO thin membranes versus the temperature is shown in Fig. 3e.

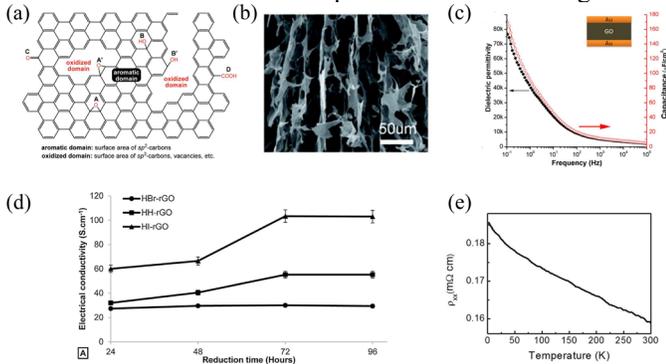

Fig. 3. Basic properties of graphene derivatives. (a) Schematic illustration of oxygen-containing groups in GO. (b) The SEM image of lateral cross section of GO sponge. (c) The relative permittivity of GO membrane. (d) Comparison of conductivity of reduced graphene oxide with different reducing agents. (e) Temperature dependence of longitudinal resistivity of the 3000-K-reduced rGO membrane. Adapted from [97], [98], [99], [110] and [112] under a Creative Commons license.

In terms of the underlying mechanism, the removal of oxygen functional groups on the surface of graphene oxide causes the carbon atoms in the GO layer to gradually revert to the $sp^2$ hybridized state, which enables the π electrons between the carbon atoms to reform a larger conjugated system, and therefore the π electrons in the conjugated system are able to move freely, which significantly improves the electrical conductivity of rGO (200-35100 S m$^{-1}$)[113], [114]. Additionally, reduced graphene oxide has a high Young's modulus (~ 200 GPa) and breaking strength (~ 115 MPa), which indicates that the rGO membrane maintains its structural integrity when it is subjected to pressure or is stretched, and is able to withstand a large amount of stress without breaking. This further improves the mechanical stability and durability of the flexible pressure sensors as well as enhances the sensitivity and detection range of the pressure sensors[115].

## III. GRAPHENE-BASED FLEXIBLE PRESSURE SENSORS

Graphene-based flexible pressure sensors can be mainly classified into piezoresistive, capacitive, and piezoelectric according to sensing mechanisms. This section mainly introduces the research progress of flexible pressure sensors based on different sensing mechanisms, and summarizes their advantages and disadvantages. The performance comparison of various flexible graphene-based pressure sensors is summarized in TABLE II.

### A. Piezoresistive Flexible Pressure Sensors

Compared to other sensing mechanisms of pressure, piezoresistive flexible pressure sensors feature excellent performance, ease of fabrication, simplicity of design, direct measurement, and low power consumption. The piezoresistive sensing mechanism is that when pressure is applied to the sensor, the deformation and strain of the sensitive membrane change thereby resulting in the change of the resistance of the sensitive membrane due to the piezoresistive effect. By detecting the change of the resistance, the amount of pressure can thus be measured and monitored.

Sensitivity and measuring range are often the two most critical parameters of a pressure sensor, where sensitivity is usually expressed as:

$$\text{The sensitivity} = R-R_0/P \tag{1}$$

where R and $R_0$ are the initial resistance and resistance under the applied pressure, respectively, P is the value of the pressure applied to the sensor.

So far, there have been many reports on flexible pressure sensors, including realizing the measurement of tiny pressure and intense pressure. However, pressure sensors capable of sensing both tiny and intense pressures have been rarely reported. In 2019, Ge et al. prepared a flexible pressure sensor (RGPS) based on rGO/polyaniline-coated sponge by combining rGO with a sponge and synthesizing polyaniline nanowires (PANI NWs) in-situ[116], and the schematic fabrication process is shown in Fig. 4a. The sensitivity of this sensor is up to 0.152 kPa$^{-1}$, and its improved sensitivity can be attributed to the formation of the hierarchically micro-protruding pimplings and air gaps within the element. This pressure sensor can monitor both small physical activities (speech recognition, swallowing, light/heavy blowing, etc.) and robust body movements (finger bending, elbow, and knee movements), and it has a relatively wide operating range (0-27 kPa) and a high SNR. The results show that the pressure sensor is highly reliable for low-pressure detection. Despite the high sensitivity of this pressure sensor, it still does not enable measurements in higher operating ranges above 27 kPa. For instance, fingertip tactile sensors call for the pressure above 100 kPa, which most of the currently reported flexible pressure sensors are unable to satisfy. Therefore, it is essential to develop flexible pressure sensors with ultra-wide operating range to meet the demand of mimicking the fingertip pressure sensing function. In 2019, Yue et al. proposed a small-size, light-weight but high-performance graphene-based pressure sensor by utilizing the reducing agent vitamin C to reduce graphene oxide, and the obtained graphene film has an advanced structure with surface fluctuation and a fluffy laminate structure in the cross-section, as shown in Fig. 4b[118]. This pressure sensor shows extraordinary performance, as depicted in Fig. 4c, with an ultra-high sensitivity of up to 10.39 kPa$^{-1}$ in a small pressure range of 0-2 kPa, an ultra-wide operating range of up to 200 kPa, a good stability and repeatability, a high operating frequency, and a fast response time and recovery time. Although the pressure sensor has a very high sensitivity in the low-pressure range, the sensitivity in the high-pressure range (> 2 kPa) is less satisfactory, with a sensitivity of 0.0034 kPa$^{-1}$, which is lower than most of pressure sensors. Moreover, the pressure sensor can only show high sensitivity in the measurement range between 0-2 kPa, which is relatively narrow for the pressure measurement range of human daily activities.

TABLE II
COMPARISON OF THE PERFORMANCE OF FLEXIBLE PRESSURE SENSORS WITH DIFFERENT SENSING MECHANISMS.

| Mechanism | Sensing materials | Sensitivity | Measuring range | Response time | Recovery time | Stability (cycles) | Year | Ref. |
|---|---|---|---|---|---|---|---|---|
| Piezoresistive | Graphene | 17.2 kPa$^{-1}$/0.1 kPa$^{-1}$ | 0 – 2 kPa/2 – 20 kPa | 120 ms | 60 ms | 300 | 2017 | [117] |
| Piezoresistive | Graphene Film | 10.39 kPa$^{-1}$/0.0034 kPa$^{-1}$ | 0 – 2 kPa/2 – 200 kPa | 11.6 ms | 25.6 ms | 1100 | 2019 | [118] |
| Piezoresistive | rGO/RGPS | 0.042 - 0.152 kPa$^{-1}$ | 0 – 27 kPa | 96 ms | - | 9000 | 2019 | [116] |
| Piezoresistive | rGOs | 5.77 kPa$^{-1}$ | 0 – 9.8 kPa | 97 ms | 98 ms | 300 | 2020 | [119] |
| Piezoresistive | Graphene-silk | 0.4 kPa$^{-1}$ | 0 – 140 kPa | - | - | 60 | 2017 | [120] |
| Piezoresistive | GMs | 52.36 kPa$^{-1}$ | 0 – 50 kPa | 100 ms | 80 ms | 6000 | 2020 | [121] |
| Piezoresistive | WGF/PVA nanowires | 28.34 kPa$^{-1}$ | 0 – 14 kPa | 87 ms | - | 6000 | 2018 | [122] |
| Piezoresistive | Alginate/graphene hydrogel | 5 kPa$^{-1}$ | 0 – 1000 kPa | 8 ms | 30 ms | 4000 | 2021 | [123] |
| Piezoresistive | rGO paper | 1.5 kPa$^{-1}$ | 0 – 60 kPa | 94 ms | 270 ms | 1200 | 2022 | [124] |
| Piezoresistive | (rGO)-cloth | 30.3 kPa$^{-1}$/0.56 kPa$^{-1}$ | 0 – 2.5 kPa/2.5 – 20 kPa | 19 ms | 23 ms | 4000 | 2022 | [125] |
| Piezoresistive | rGO-PP composite | 0.46 - 3.05 kPa$^{-1}$ | 24 – 122 kPa | 16 ms | - | 7000 | 2021 | [126] |
| Piezoresistive | rGO/PPyF | 32.32 kPa$^{-1}$ | 0 – 21 Pa | - | - | 1000 | 2021 | [127] |
| Piezoresistive | MX/rGO PET | 0.71 kPa$^{-1}$/0.07 kPa$^{-1}$ | 0 – 11 kPa/11 – 100 kPa | 134 ms | 136 ms | 1500 | 2022 | [128] |
| Piezoresistive | rGO-PDMS/SR | 7.0838 kPa$^{-1}$/ 0.0168 kPa$^{-1}$ | 0 – 50 kPa/50 – 250 kPa | - | - | 200 | 2023 | [129] |
| Capacitive | Graphene oxide foam | 0.8 kPa$^{-1}$ | 0 – 1 kPa | ~ 100 ms | ~ 100 ms | 1000 | 2017 | [130] |
| Capacitive | Graphene | 0.33 kPa$^{-1}$ | < 1 kPa | < 20 ms | - | 1000 | 2018 | [131] |
| Capacitive | GR/PDMS sponge | 0.12 kPa$^{-1}$ | 0 – 500 kPa | ~ 7 ms | 60 ms | 5000 | 2019 | [132] |
| Capacitive | borophene-graphene aerogel | 0.90 kPa$^{-1}$ | < 3 kPa | ~ 110 ms | 180 ms | 1000 | 2021 | [133] |
| Capacitive | LIG and porous PDMS | 0.011 kPa$^{-1}$/0.026 kPa$^{-1}$/0.004 kPa$^{-1}$ | 0 – 15 kPa/15 – 40 kPa/40 – 100 kPa | ~ 120 ms | - | 5000 | 2021 | [134] |
| Capacitive | porous foams of PDMS | 3.7 kPa$^{-1}$ | 2 – 6 kPa | - | - | - | 2022 | [135] |
| Capacitive | LIG/composite dielectric structure | 2.52 kPa$^{-1}$ | 0 – 10.4 kPa | ~ 39 ms | ~ 48 ms | 1000 | 2023 | [136] |
| Capacitive | RGO cotton fiber | 15.84 kPa$^{-1}$ | 0 – 500 kPa | - | - | 400 | 2023 | [137] |
| Capacitive | GO/DA/ PANI | - | 0 – 25.48 kPa | - | - | 150 | 2023 | [138] |
| Piezoelectric | PTNWs/Graphene | 9.4 × 10$^{-3}$ kPa$^{-1}$ | - | 5 – 7 ms | - | - | 2017 | [139] |
| Piezoelectric | Graphene/ZnO | 1.7 nA kPa$^{-1}$ | 0 – 100 kPa | ~ 3 s | ~3 s | - | 2020 | [140] |
| Piezoelectric | PVDF/GO | 38.8 mV/N | 20 – 70 kPa | - | - | - | 2022 | [141] |

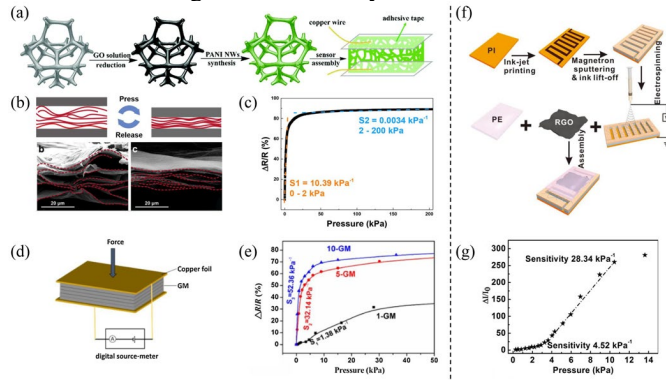

Fig. 4. Piezoresistive flexible pressure sensors. (a) Fabrication process of the flexible pressure sensor. (b) Schematic diagrams of the states of a graphene film in a flexible pressure sensor with pressure and without pressure. (c) The resistance change rate of a graphene based flexible pressure sensor in response to pressure. (d) Structure diagram of a pressure sensor based on N-layer graphene. (e) Resistance changes with pressure for GM sensor with 1, 5, and 10 atomic layers of graphene films. (f) The schematic of the fabrication of the flexible piezoresistive sensor. (g) ΔI/I$_0$ of the pressure sensors versus ΔP on it. Adapted from [116], [118], [121] and [122] under a Creative Commons license.

Liu et al. proposed a flexible pressure sensor based on a layer-by-layer structure[121]. Specifically, it was prepared by stacking different layers of graphene films (GM) and reduced graphene oxide films (rGOM) as sensing materials, respectively, where the structure of the GM-based pressure sensor is schematically shown in Fig. 4d. Compared to the pressure sensor described above[118], this sensor features ultra-high sensitivity in the pressure range of 0-5 kPa, with a sensitivity as high as 52.36 kPa$^{-1}$. As illustrated in Fig. 4e, it is attributed to the fact that the more the number of GM layers, the more the number of air gaps inside the pressure sensor. And as a slight pressure is applied to the sensor, the contact area between the layers is significantly increased and thereby the sensitivity of the pressure sensor is improved. However, in the range of 5-20 kPa, the interlayer structure becomes compact, and the rate of change of the interlayer contact area is greatly reduced, resulting in a decrease of sensitivity. Hence, achieving an optimal balance between high sensitivity and a wide pressure detection range remains a technical challenge.

There are two main factors that affect the performance of piezoresistive pressure sensors: one is the sensitive materials used to sense the pressure, and the other is the structure of the sensor that can be used to adjust the contact resistance between the sensing materials and the electrodes on the substrate. Liu et al. proposed a method of fabricating a pressure sensor that formats the interconnected polyvinyl alcohol (PVA) nanowires between the fabricated interdigital electrodes (IDE) and a piece of ultrathin wrinkled graphene film (WGF)[122]. The formation process of this pressure sensor is shown in Fig. 4f. Due to the synergistic effect of WGF and embedded PVA nanowires, there was a significant increase in the contact area and the conductive paths, resulting in improved sensitivity of the pressure sensor. The pressure sensor has a sensitivity of 28.34 kPa$^{-1}$ in the high-pressure range of over 3 kPa, as depicted in Fig. 4g. Moreover, they investigated the minimum detection limit of the pressure device to further assess the detective capability. The results show that the pressure sensor has excellent reliability for low pressure detection.

Human skin is an ultra-sensitive, flexible, and stretchable sensing unit with a wrinkled structure, and researchers have also been working on developing a touch device that mimics the natural sensing network of human skin. In 2019, Jia et al. reported a novel graphene-based pressure sensor inspired by wrinkles on the surface of human skin[142]. With a sensitivity of up to 178 kPa$^{-1}$, this sensor can also monitor pressures as small as 42 Pa. Although the performance exceeds that of the vast majority of previously reported piezoresistive pressure sensors, it still fails to meet the requirements for the detection of lower pressure ranges. Weng et al. obtained wrinkled graphene oxide structures (wrGOs) by a simple stretch-release method, and fabricated the flexible pressure sensor based on wrGOs/tape composites[119]. The results show that the sensitivity of the pressure sensor in the low-pressure range (0-490 Pa) is 5.77 kPa$^{-1}$. Although it has a high sensitivity in the tiny pressure range to detect low pressure, the sensitivity in the large range (490 Pa - 9.8 kPa) is only 0.25 kPa$^{-1}$.

Apart from research above on pressure sensors based on pleated structures, some researchers work on the implementation of pleated structures on pressure sensors using other more straightforward methods, which specifically involve stretching and releasing the sample to achieve pleating, which can be realized on a large scale due to its simplicity of fabrication and low cost. Thermal wrinkling is a way to generate wrinkles, specifically by creating a difference in the coefficient of thermal expansion between different structures, which produces thermal residual stresses that lead to the formation of wrinkles. Ge et al. prepared a flexible pressure sensor with a pleated structure by rapid thermal annealing of spin-coated PEDOT: PSS (PH1000) on graphene-PDMS composites to promote the generation of pleat[143]. The sensitivity of the sensor was 2.83 kPa$^{-1}$ in the low-pressure range of 0-490 Pa and 0.12 kPa$^{-1}$ in the range of 490 Pa-10 kPa. However, the sensitivity of the sensors in the high-pressure region is much lower than that in the low-pressure range, adversely affecting the ability to monitor high-pressure ranges.

In addition to the two important metrics of sensitivity and pressure range, linearity is also a vital indicator of the accuracy of pressure measurements. Pang et al. proposed a graphene pressure sensor with a random, highly distributed spine-like structure using abrasive paper as a template by thermal reduction method, which exhibits a high sensitivity of up to 25.1 kPa$^{-1}$ over a wide pressure range of 0-2.6 kPa with good linearity[144]. Compared with the pressure sensors that were previously reported[116], [117], [118], [119], this pressure sensor has both high sensitivity and excellent linearity.

### B. Capacitive Flexible Pressure Sensors

Compared to resistive pressure sensors, capacitive pressure sensors may have higher sensitivity and can detect tinier pressure changes because changes in capacitance are typically more sensitive than changes in resistance. Moreover, capacitive pressure sensors also have the advantages of good stability, low power consumption, and large fabrication area, and have received extensive attention from researchers[14], [145]. Typically, a capacitive sensing structure consists of two conductive electrodes as well as an intermediate dielectric layer, and then the capacitance (C) is given by the following equation:

$$C = \varepsilon_0 \varepsilon_r A / d \quad (2)$$

where $\varepsilon_0$ is the vacuum dielectric constant, $\varepsilon_r$ is the relative permittivity, A is the overlap area of the two conductive plates and d is the distance between the pole plates. According to the above equation, the primary sensing mechanism of capacitive sensors is that when pressure is applied, the capacitance size will change due to the distance or area between the two plates or dielectric constant, and the change in pressure can be calculated by detecting the change in capacitance. The sensitivity of the capacitive pressure sensor can be expressed as:

$$S = (\Delta C / C_0) / \Delta P \quad (3)$$

where $\Delta C$ is the change in capacitance, $C_0$ is the initial capacitance, $\Delta P$ is the value of the change in pressure applied to the sensor.

Typically, most of the flexible capacitive pressure sensors use the change of capacitive dielectric layer to detect the pressure magnitude, thus the magnitude of the relative dielectric constant of the dielectric layer is vitally crucial for the sensitivity of the pressure sensor. Graphene oxide exhibits a high relative dielectric constant, and many studies have shown that the foam graphene oxide has high dielectric constant and good elasticity properties, and is an ideal dielectric layer for fabricating flexible pressure sensors with high-sensitivity. Wan et al. fabricated a flexible pressure sensor based on a graphene oxide foam structure by preparing reduced graphene oxide patterns on a PET substrate as the upper and lower electrode plates and using the graphene oxide foam as a dielectric layer for capacitive pressure sensors[130]. The graphene oxide foam has porous microstructures, and these porous structures provide a void for the elastic deformation of the graphene oxide foam when subjected to external pressure, which provides an effective space for the change of capacitance.

For capacitive flexible pressure sensors, in addition to selecting favorable dielectric materials, highly flexible and highly conductive electrode structures are also crucial for improving the performance of the pressure sensors. In 2014, Lin et al. fabricated three-dimensional porous graphene foams, known as laser-induced graphene(LIG), by inscribing a $CO_2$ infrared laser on a polyimide (PI) thin film[146]. In recent years, laser-induced graphene has been used as an electrode structure for high-performance flexible pressure sensors due to its advantages of high flexibility, high stretchability, and high conductivity. Huang et al. proposed a capacitive flexible pressure sensor consisting of laser-induced graphene and polydimethylsiloxane (PDMS) foam[134], using LIG as the electrode structure and porous PDMS foam as the dielectric layer. The fabrication process is shown in Fig. 5a. The citric acid monohydrate (CAM) was used as a pore-forming agent to fabricate the porous PDMS foam structure to improve the sensitivity of the pressure sensor. In contrast, the porosity of the foam was determined by the ratio of CAMs to PDMS. Subsequently, in order to study the relationship between sensitivity and the porosity and thickness of CAMs/PDMS, respectively, the researchers prepared four flexible capacitive pressure sensors with different ratios of CAMs to PDMS (CAMs: PDMS = 0: 1, 1: 1, 3: 2, and 2: 1, all of which had a thickness of 1.5 mm) and different thicknesses (1.5 mm, 3 mm, 4.5 mm, and 6 mm, CAMs: PDMS = 2: 1). The results of the two sets of control experiments (Fig. 5b-c) clearly show that the sensitivity of flexible capacitive pressure sensor (FCPS) with PDMS foam is more than five times higher than that of FCPS without PDMS foam, demonstrating the superiority of PDMS foam in high-performance pressure sensors. Moreover, the capacitive response of the FCPS with a smaller thickness of CAMs/PDMS has a larger variation and higher sensitivity. This is attributed to the fact that the thinner thickness of CAMs/PDMS allows for a larger deformation of the electrode spacing under a certain pressure load, resulting in higher elasticity.

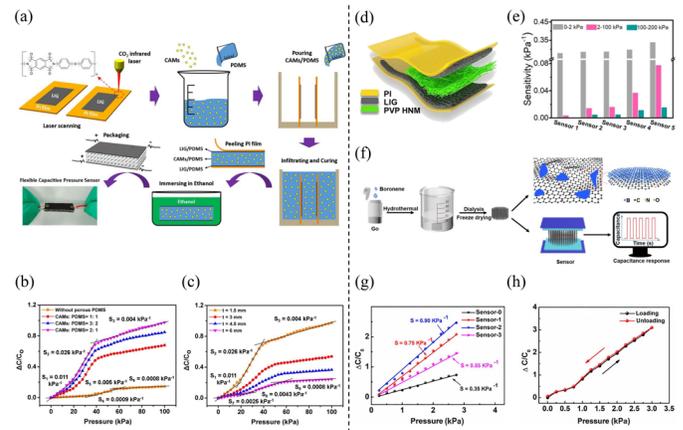

Fig. 5. Capacitive flexible pressure sensors. (a) Fabrication process of the flexible capacitive pressure sensor. (b) Pressure response of the flexible capacitive pressure sensor with different ratios of CAMS:PDMS. (c) Pressure response of the flexible capacitive pressure sensor with different thicknesses. (d) Schematic structure of the flexible capacitive pressure sensor. (e) Sensitivities of the five sensors at different pressure ranges. (f) The schematic of the fabrication process of the flexible capacitive pressure sensor. (g) The comparison of the sensitivity of sensors with different borophene content. (h) Capacitive response behaviors when the pressure is continuously applied and removed. Adapted from [134], [147] and [133] under a Creative Commons license.

Apart from using of PDMS foam to form a porous structure, other methods can be used to form a porous structure with a dielectric layer. In 2023, Guo et al. prepared a flexible capacitive pressure sensor using polyvinylpyrrolidone (PVP) layered nanofibrous films (HNMs) as a dielectric layer and laser-induced graphene as an electrode layer[147]. The schematic structure of the pressure sensor is indicated in Fig. 5d. By designing five flexible pressure sensors with different combinations of dielectric materials and electrode structures and measuring their sensitivities in different pressure ranges respectively, the measurement results show that the pressure sensor consisting of LIG electrodes and PVP HNM dielectric layer has the highest sensitivity value (0.338 kPa$^{-1}$) in the pressure range of 0-2 kPa (Fig. 5e). And in the pressure range of 2-100 kPa, its sensitivity value (0.078 kPa$^{-1}$) is 22 times, 5 times, 4.7 times and 2 times higher than the other four sensors. This shows that optimization of the dielectric layer and electrode structure plays a pivotal role in improving the device's performance such as sensitivity.

In general, combining two 2D materials improves the electrical conductivity, mechanical properties, and porosity of the material[148], [149], which can enhance the sensitivity of flexible capacitive pressure sensors. In 2021, Long et al. prepared a capacitive pressure sensor with a borophene-graphene aerogel as the dielectric layer by doping graphene oxide (GO) with borophene[133]. The preparation process is shown in Fig. 5f, which compares the performance of pressure sensors with different borophene additions. The sensitivity of pressure sensors using different aerogels as the dielectric layer is illustrated in Fig. 5g, and the results show that pressure sensors containing a moderate dose of borophene exhibit the highest sensitivity (0.9 kPa$^{-1}$). There are generally two reasons for the increased sensitivity. On the one hand, more defect states are introduced with the increase of borophene content and the addition of these defects enhances its charge storage capacity, which can lead to more tremendous capacitance changes under pressure. On the other hand, the combination of borophene and graphene increases the porosity and specific surface area of the aerogel, resulting in a greater change in capacitance under pressure. Additionally, the sensor has a good recovery characteristics and high repeatability (Fig. 5h).

## C. Graphene Field-effect Transistor Flexible Pressure Sensors

In recent years, the development of field-effect transistor (FET) based pressure sensors has received a widespread attention from researchers, which can be attributed to their remarkable properties, including low power consumption, high spatial resolution, low noise, easy integration with circuits, and avoidance of crosstalk signals. Furthermore, for application scenarios such as gesture recognition, which is common in life and requires a large amount of data to be acquired on arrays rather than single sensors, FET based pressure sensors are well-suited to meet this requirement due to their utilization of the advantages of active matrix sensing arrays. Specifically, a FET is a transistor that controls the current by controlling the electric field inside the device, which consists of three main parts: the source, the drain, and the gate, and controls the number of charge carriers in the channel region between the source and drain by changing the gate voltage, thereby altering the current ($I_d$) between the source and drain. For example, Sun et al. developed a pressure sensor array based on coplanar gate graphene FETs by laminating a top cover with a graphene square pattern onto a graphene field-effect transistor (GFET) backplane film with a pressure-sensitive element [150] (Fig. 6a). The design of coplanar gate GFETs and the use of graphene as the electrodes and the conductive channel simplified the fabrication process of the sensor. As shown in Fig. 6b, the sensor exhibits a high pressure sensitivity of 0.12 kPa$^{-1}$ and a mechanical durability of more than 2500 cycles. In addition, the sensor can operate at a low voltage of less than 2 V due to the use of a highly capacitive ion gel as the gate dielectric material, which reduces the energy consumption of the sensor system.

However, the use of an ion gel gate dielectric limits pressure detection at a higher range, and in order to achieve detection at a wider pressure range, Sung-ho shin et al. fabricated a graphene FET pressure sensor with an air dielectric layer by folding an origami substrate consisting of two plastic panels and a foldable elastic connector [151] (Fig. 6c). When pressure is applied to the sensor, the thickness of the air gaps and elastomeric partition spacers decreases with the increase of the applied pressure, leading to an increase in the capacitance of the metal-air-graphene structure, which causes the gate voltage ($V_g$) to sense more charge carriers in the graphene channel, and thus detect the change in pressure through the change in current. By utilizing the air gap formed by the folded substrate as the dielectric layer of the sensor and maintaining a clean interface between the graphene channels and air, the pressure sensor achieves high sensitivity and a wide pressure detection range (250 Pa ~ 3 MPa). In addition, this simple fabrication method eliminates the need for additional pressure-sensing components or layers and allows for the formation of high-density arrays of integrated pressure sensors (Fig. 6d), which is essential for improving the spatial resolution of pressure sensors and achieving higher accuracy in tactile sensing.

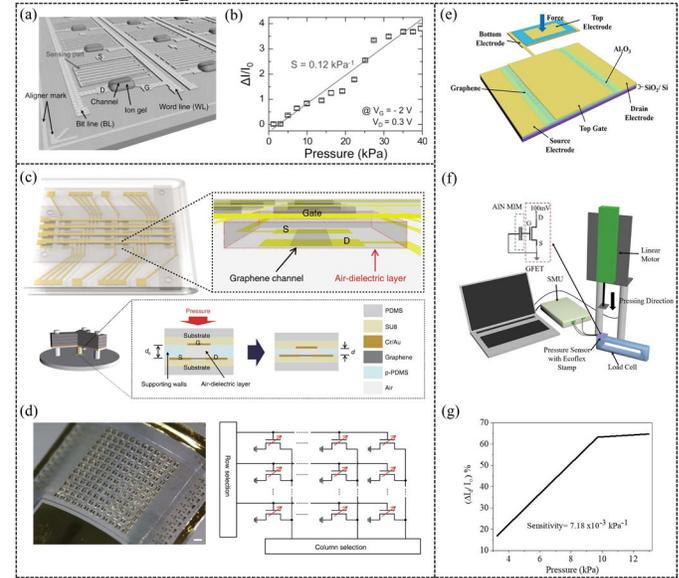

Fig. 6. Graphene field-effect transistor flexible pressure sensors. (a) Schematic diagram of the pressure sensor based on coplanar gate GFETs with an ion gel gate dielectric. (b) Sensitivity of the GFET pressure sensor. (c) Schematic diagram and sensing mechanism diagram of a folded pressure-sensitive graphene FET with an air dielectric layer. (d) 12 × 12 active-matrix pressure-sensitive FET array and its electronic circuitry schematic. (e) Scheme of the pressure sensor with AlN capacitor connected to GFET in an extended gate configuration. (f) Experimental device diagram of FET pressure sensor. (g) Sensitivity of graphene field-effect transistor pressure sensor. Adapted from [150], [151] and [152] under a Creative Commons license.

Presently, piezoelectric materials are widely used in the extended gate structure of field effect transistor pressure sensors, whereas the high voltage required during the polarization of piezoelectric materials may cause damage to the transistor device and also cause high power consumption problem. To solve this issue, Nivasan Yogeswaran et al. proposed a GFET pressure sensor coupled by a GFET and an Aluminium Nitride (AlN) piezoelectric capacitor [152] (Fig. 6e). Due to the good crystal orientation structure, the AlN material does not require polarization, which allows the pressure sensor to operate at low voltages. At the same time, the piezoelectric potential regulates the voltage at the gate electrode of the GFET, which in turn regulates the value of the channel current. Combined with the high carrier mobility and intrinsic mechanical properties of graphene itself, the pressure sensor achieves ultra-low operating voltage (~100 mV) as well as high sensitivity detection of the pressure (7.18 × 10$^{-3}$ kPa$^{-1}$), and the detection device and sensitivity curves are shown in Fig. 6f-g.

## D. Other Types of Flexible Pressure Sensors

Apart from the above dominant sensing mechanisms, researchers have also developed pressure sensors with other sensing mechanisms, such as piezoelectric. Firstly, although graphene itself does not have a piezoelectric effect, it can be combined with other piezoelectric materials to form higher performance pressure sensors. For example, Chen et al.

presented a pressure sensor based on $PbTiO_3$ nanowires(PTNWs)/graphene heterostructures[139], and the manufacturing process of this pressure sensor is shown in Fig. 7a. Unlike conventional pressure sensors, this pressure sensor enables static pressure measurements. The pressure responses of three pressure sensors (pure PTNWs-based, graphene-based, and PTNWs/graphene-based) under pressure pulses are compared in Fig. 7b-d. The results show that the pure PTNWs-based pressure sensors have only transient responses and do not record sustained responses when the applied pressure is kept constant, as shown in Fig. 7b. Moreover, since the lattice of graphene does not produce significant changes at smaller pressures, the graphene-based pressure sensors also do not have current changes, as shown in Fig. 7c. When graphene and PTNWs are combined to form heterostructures of PTNWs/graphene, the strain-induced polarization charge generated by PTNWs increases the scattering source of graphene, which reduces the carrier mobility of graphene. Therefore, when the external pressure remains constant, the polarization charge distribution on the PTNWs also remains constant. The carrier mobility in graphene will continue to decrease, resulting in a continuous and stable current change and thereby achieving the measurement of static pressure (Fig. 7d). This is of great significance for the application of the pressure sensors in the fields of medical and health monitoring, and human movement monitoring. Furthermore, the synergistic effect of the two materials enabled the pressure sensor to exhibit a sensitivity of up to $9.4 \times 10^{-3}$ $kPa^{-1}$ and a fast response time of 5-7 ms. Likewise, taking advantage of the synergistic effect between piezoelectric materials and graphene, Li et al. prepared a flexible pressure sensor based on the heterostructure of graphene and PZT nanowires[153], and the preparation process is shown in Fig. 7e. As depicted in Fig. 7f, it showed a current change of 65 μA as the pressure is applied, which is much higher than that (~12 μA) of the pure graphene-based pressure sensors under the same pressure. This can be attributed to the piezoelectric properties of the PZT nanowires resulting in an increase in carrier scattering of graphene, and thus an increase in the pressure sensor's response to the pressure change.

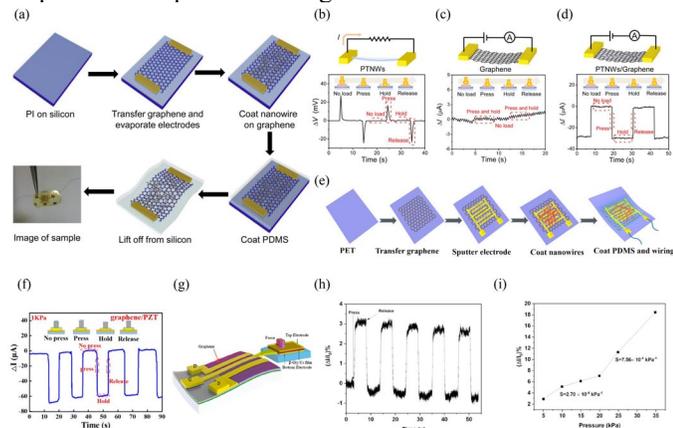

Fig. 7. Other types of flexible pressure sensors. (a) Fabrication process of the PTNWs/G pressure sensor. (b) Pressure response of a pure PTNW-based pressure sensor under a pressure pulse. (c) Pressure response of a graphene-based pressure sensor under a pressure pulse. (d) Pressure response of a PTNWs/G pressure sensor under a pressure pulse. (e) Schematic diagram of the fabrication process of the PZT nanowires/G pressure sensor. (f) Current response of the sensor under a pressure of 1 kPa. (g) Schematic diagram of the pressure sensor structure. (h) Dynamic response of the sensor at a pressure of 5 kPa. (i) Sensitivity of pressure sensor under different pressure ranges. Adapted from [139], [153] and [154] under a Creative Commons license.

Besides the piezoelectric sensing mechanism mentioned above, in 2020, Nivasan Yogeswaran et al. proposed to combine graphene-based field effect transistors with piezoelectric materials to form a novel pressure sensor[154]. The structure of this sensor is illustrated in Fig. 7g. Specifically, when the external pressure is applied to the piezoelectric layer, the resulting piezoelectric potential difference affects the gate voltage of the GFET, which in turn modulates the electron density in the graphene channel and enables the detection of pressure changes. Due to the extended gate configuration of the sensor and direct connection of the piezoelectric sensing layer to the gate of the GFET, the requirement for the operating voltage is reduced, allowing the pressure sensor to operate at very low voltages (50 mV). Moreover, due to the high electron mobility property of graphene and the high sensitivity of piezoelectric materials, the sensitivity of this pressure sensor can reach up to $2.7 \times 10^{-4}$ $kPa^{-1}$ in the range of 5-20 kPa, and up to $7.56 \times 10^{-4}$ $kPa^{-1}$ in the range of 20-35 kPa, as shown in Fig. 7h-i.

## IV. SUSPENDED GRAPHENE-BASED GAS PRESSURE SENSORS

For pressure sensors based on suspended thin membranes, there are several advantages: (1) The suspended structure avoids problems such as carrier scattering and phonon leakage caused by contact between graphene and the substrate and helps to maintain the thermal stability properties and electronic properties of graphene; (2) The graphene membranes as a sealing membrane were suspended over the cavity for gas pressure sensors, which leads to the formation of a change in the pressure difference between the interior and the exterior of the cavity as the surrounding pressure is changed. In terms of sensing mechanisms, pressure sensors based on suspended graphene membranes are mainly classified as piezoresistive, capacitive, and resonant. In this section, we discuss the performance of pressure sensors based on suspended graphene membranes under different sensing mechanisms. The performance comparison of various graphene-based pressure sensors is summarized in TABLE III.

### A. Graphene-Based Piezoresistive Pressure Sensors

Piezoresistive effect is the most widely used among the various sensing mechanisms because of its large dynamic range, low complexity, small size, high reliability, and low cost. In piezoresistive pressure sensors, the applied pressure acts on the diaphragm, resulting in the deformation and stress and strain and consequent resistance change in the diaphragm.

In 2013, Anderson D. Smith et al. developed the first pressure sensor based on a suspended monolayer graphene membrane. The sensor works by detecting pressure changes through strain-induced changes in the band structure of the graphene band(Fig. 8a)[96]. The results show that the sensor with suspended graphene membrane shows a high sensitivity to the change in resistance when subjected to pressure, thus demonstrating that mechanical bending and stretching of the suspended graphene membrane leads to a strong response property of its resistance to pressure. However, the suspended graphene membrane of this pressure sensor does not have a support layer, which primarily causes problems such as fracture of the graphene membrane. Thus, in 2016, Wang et al. proposed a graphene pressure sensor and the structural schematic of the pressure sensor is shown in Fig. 8b, which was realized by transferring a large-area, few-layered graphene onto a suspended silicon nitride membrane perforated by a periodic array of micro-through-holes[169]. Each through-hole is covered with a circular drum-like graphene layer, namely a graphene "microdrum". The innovation of this pressure sensor design is the introduction of an array of through-holes into the supporting nitride membrane so that when pressure is applied, the graphene membrane can yield greater strain due to the localized deformation of the holes. Moreover, the deflection of the perforated film is greater compared to the previous imperforated film[155]. Measurement results indicate that the gauge factor of the graphene membrane is 4.4, and the sensitivity of this pressure sensor is $2.8 \times 10^{-5}$ mbar$^{-1}$, which is higher than that of the pressure sensor that is based on graphene membrane alone ($2.96 \times 10^{-6}$ mbar$^{-1}$). However, when fabricating graphene-based pressure sensors, it is usually necessary to protect the graphene membranes, which prevents other environmental factors (e.g., temperature, humidity) from interfering with the sensing membranes. Moreover, the performance of graphene devices depends significantly on the quality of the graphene, and if not adequately protected, the graphene is susceptible to oxidation and contamination, which will greatly affect the conductivity of the graphene, and thus essential properties such as the sensitivity of the pressure sensor.

TABLE III
PERFORMANCE COMPARISON OF SUSPENDED GRAPHENE PRESSURE SENSORS.

| Mechanism | Material | Sensitivity | Pressure range | Dimensions($\mu$m) | Year | Ref. |
|---|---|---|---|---|---|---|
| Piezoresistive | Graphene | $2.96 \times 10^{-5}$ kPa$^{-1}$ | 20 kPa – 100 kPa | 6 × 64 | 2013 | [96] |
| Piezoresistive | Graphene/SiN$_x$ | $8.5 \times 10^{-5}$ kPa$^{-1}$ | 0 – 70 kPa | 280 × 280 | 2013 | [155] |
| Piezoresistive | Graphene/Perforated SiN$_x$ film | $2.8 \times 10^{-5}$ mbar$^{-1}$ | 0 – 40 kPa | 490 × 490 | 2016 | [156] |
| Piezoresistive | BN-graphene-BN | $1.87 \times 10^{-4}$ kPa$^{-1}$ | 100 – 200 kPa | 6 × 64 | 2018 | [157] |
| Piezoresistive | PMMA/Graphene | $2.87 \times 10^{-5}$ kPa$^{-1}$ | 0 – 80 kPa | r=5 | 2019 | [158] |
| Piezoresistive | PMMA/Graphene | $7.42 \times 10^{-5}$ kPa$^{-1}$ | 0 – 70 kPa | r=8 | 2020 | [159] |
| Piezoresistive | h-BN/Graphene/h-BN | $2.9 \times 10^{-4}$ kPa$^{-1}$ | -80 – 0 kPa | 210 × 210 | 2022 | [160] |
| Piezoresistive | Monolayer Graphene | $4.03 \times 10^{-5}$ kPa$^{-1}$ | 0 – 500 Pa | 1400 × 1400 | 2022 | [161] |
| Piezoresistive | Si$_3$N$_4$/Graphene | $5.32 \times 10^{-4}$ kPa$^{-1}$ | 10 Pa – 100 kPa | 64 × 9 | 2023 | [162] |
| Capacitive | Graphene | 15.15 aF Pa$^{-1}$ | 0 – 1800 Pa | r=750 | 2016 | [163] |
| Capacitive | Graphene polymer | 123 aF Pa$^{-1}$ mm$^{-2}$ | 100 kPa | r=15 | 2017 | [164] |
| Capacitive | DL Graphene | 47.8 aF Pa$^{-1}$ mm$^{-2}$ | -10 – 0 kPa | r=2.5 | 2020 | [165] |
| Capacitive | Graphene/PMMA | $-47.5$ dB V (4.22 mV/Pa) | 65 – 138 kPa | r=1750 | 2021 | [166] |
| Resonant | Graphene | 9000 Hz/mbar | 0.8 – 100 kPa | r=2.5 | 2016 | [167] |
| Resonant | Au/Graphene | $-0.70$ kHz/min | $7 \times 10^{-7}$ – 100 kPa | r=20 | 2022 | [168] |

It has been demonstrated that some polymers such as PMMA, parylene-C, polydimethyl siloxane, polyethylene glycol terephthalate, and poly (4-vinylphenol) not only retain the high electrical conductivity and mechanical properties of graphene[170], [171], but also ensure that the integrity of the graphene is not damaged, which can greatly increase the yield of the device. These are attributed to a number of excellent properties of the polymer. Firstly, the highly controllable and uniform thickness of the polymer effectively insulates the environment from moisture and other chemical molecules, thereby increasing the conductivity of the graphene film[172]. At the same time, the polymer's high-temperature stability helps to improve the performance and stability of graphene in high-temperature applications. Secondly, the presence of polymers creates a low bending stiffness, which allows the graphene film to adhere more firmly to the substrate through van der Waals forces[173]. Furthermore, the polymer coating improves the friction and coefficient of friction of the graphene film, preventing damage to the graphene during repeated sliding[172]. In addition to some polymers, hexagonal boron nitride (h-BN) has become a popular protective material due to its excellent properties. Firstly, h-BN has an atomically smooth surface with an atomic structure similar to graphene and a small lattice mismatch with graphene (1.7%)[174], which helps to form a flat, stable heterostructure and maintains the electronic properties of graphene to the maximum extent. Secondly, h-BN has a similar coefficient of thermal expansion to graphene, which reduces stress issues due to thermal expansion mismatch. Finally, h-BN is also extremely chemically stable, and graphene-based devices can operate over an extended temperature range of up to 500°C when protected with h-BN [175]. In addition, boron nitride is also commonly used as a substrate for graphene, and it has been shown that the mobility and inhomogeneity of graphene transferred to a boron nitride substrate is almost an order of magnitude better than that of bilayer graphene on $SiO_2$[176]. Therefore, we can conclude that some crucial properties are essential when choosing a protective material for graphene, including good high-temperature stability, excellent surface flatness, and a low stress interaction coefficient with graphene. These characteristics collectively ensure that the protective material not only withstands the effects of harsh environments on graphene, but also maintains its structural and performance integrity.

In 2018, Li et al. proposed a pressure sensor based on a graphene-boron nitride (BN) heterostructure, which consists of a single layer of graphene sandwiched between two vertically stacked BN nanofilms[157], and the fabrication process is shown in Fig. 8c. The boron nitride layer was used to protect the graphene layer from oxidation and contamination, and the characteristics of the sensor were investigated experimentally with a sensitivity of up to 24.85 μV/V/mmHg over a pressure range of 130-180 kPa, but the sensitivity was lower below 130 kPa and above 180 kPa. The relative resistance of the BN-graphene-BN pressure-sensitive element changed by only 3.1% after exposure to the environment for seven days. While, graphene-based pressure sensor without BN protective layer has a relative resistance change of 15.7%, as illustrated in Fig. 8d. Thus, this method shows that the BN layer is a good passivation layer for graphene-based pressure sensors and effectively improves the sensitivity and stability of graphene-based pressure sensors.

Similarly, utilizing h-BN to protect graphene, Wang et al. prepared a highly sensitive graphene pressure sensor[160], in which the preparation process flow is shown in Fig. 8e-f. Its sensing element is completely isolated from the external environment by the initial protection of h-BN and the secondary protection of Cu-Sn solid-liquid diffusion bonding. The sensor's sensitivity is measured to be $2.9 \times 10^{-4}$ $kPa^{-1}$ in the pressure range of -80 to 0 kPa. In addition, the relative resistance of the graphene pressure sensor changes by only 2.3% after 30 days of exposure to the environment, and the resistance remains stable even in high-temperature or high-humidity environments, suggesting that this protection mechanism effectively improves the durability and stability of the pressure sensor.

Although graphene pressure sensors have made significant progress in terms of sensitivity, there are still some challenges and difficulties in other aspects. For example, the load-bearing capacity of the suspended two-dimensional membrane limits the detection range of the pressure sensors. Therefore, on the one hand, it is necessary to introduce a support diaphragm to the graphene membrane to improve its pressure detection range. On the other hand, as mentioned earlier, since contaminants inevitably remain on the surface of the graphene membrane and are prone to react with water molecules or other gas molecules in the air, all of which can reduce the stability of the pressure sensor to some extent. Therefore, it is also crucial to select a superior encapsulation material for graphene. Silicon nitride has ultra-high mechanical strength and ultra-low losses[179], and effectively isolates graphene from oxygen and water, enhancing the stability and reliability of graphene layers that are sensitive under extreme conditions. In 2022, Zhu et al. investigated a graphene-based pressure sensor with a wide range and high repeatability[177]. The sensor device is shown in Fig. 8g. The study transferred graphene into a silicon pressure-bearing elastic diaphragm protected by $Si_3N_4$ nanofilms. The results showed a significant improvement in the measurement range and repeatability of the graphene pressure sensor, demonstrating its potential for a variety of applications such as industrial control and health monitoring. Although this study demonstrated improved stability, controlling the effects of temperature variations on the graphene pressure sensors remains challenging.

The resistance of graphene is extremely sensitive to changes in temperature, this is because changes in temperature cause significant changes in the carrier concentration of graphene, which ultimately leads to changes in the resistance of graphene. This can make it difficult to meet the requirements of high-precision pressure measurement. Although the pressure sensors above can withstand temperatures as high as 130 °C, they cannot work properly in higher temperature environments, which limits their applications in aerospace and more industrial fields. Hence, in order to improve the ability of the pressure sensors to work stably in high temperature environments, Zeng et al. proposed a novel high-temperature pressure sensor with a fabrication process depicted in Fig. 8h, which used silicon nitride ($Si_3N_4$) as the protective layer of graphene[178]. In general, the temperature-sensitive properties of graphene are mainly affected by electron-phonon coupling and the thermal expansion effect of graphene/substrate in high-temperature environments. The thermal expansion effect dominates below 350 °C, leading to an increase in the resistance of the graphene layer due to thermal stress, while the electron-phonon coupling effect changes more significantly at temperatures higher than 350 °C, leading to a decrease in the resistance of graphene. This study compared the resistance characteristics of graphene pressure sensors with and without $Si_3N_4$ protection, as illustrated in Fig. 8i. It can be seen that the resistance of $Si_3N_4$-coated pressure sensor does not change significantly with temperature, indicating that the device can maintain relatively stable performance during temperature fluctuations (up to 420 °C), which can be attributed to the excellent thermal shock properties of silicon nitride[180]. However, due to the increase in membrane thickness, the maximum deformation under the same load decreases, which leads to a decrease in the sensitivity of the sensor (from $2.33 \times 10^{-3}$ kPa$^{-1}$ to $5.32 \times 10^{-4}$ kPa$^{-1}$). Overall, this graphene pressure sensor with a protective $Si_3N_4$ layer has significant potential for application in challenging environments such as the petrochemical, automotive, and electronics industries.

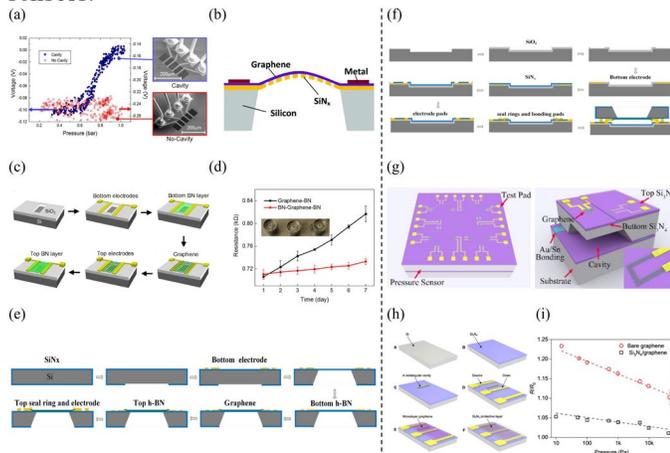

Fig. 8. Graphene-based piezoresistive gas pressure sensors. (a) Voltage-pressure response curves of the device with and without cavities. (b) Schematic diagram of a pressure sensor using a perforated array of silicon nitride as a graphene-supported membrane. (c) Schematic of the fabrication process of the BN–graphene–BN pressure sensing elements. (d) Comparison of air stability of BN-graphene-BN pressure sensing elements and BN-graphene pressure sensing elements. (e) The upper part of the device process flow. (f) The bottom part of the device process flow. (g) Schematics and sectional drawing of the pressure sensor. (h) Fabrication process of the graphene pressure sensor protected by a $Si_3N_4$ layer. (i) Comparison of piezoresistive characteristics of graphene pressure sensor with and without silicon nitride protection. Adapted from [96], [169], [157], [160], [177] and [178] under a Creative Commons license.

### B. Graphene-Based Capacitive Pressure Sensors

Compared with piezoresistive pressure sensors, capacitive pressure sensors are widely used in pharmaceutical, aerospace, and automotive applications due to their high sensitivity, low-temperature dependence, high signal response, and low power consumption.

In the capacitive graphene pressure sensors, the graphene membrane is generally used as the upper electrode plate of the capacitor. When there is a pressure difference between the sealed cavity and the outside atmosphere, the graphene membrane will deform up and down. That is, the deformation of the graphene membrane changes, which leads to changes in the distance between the graphene membrane and bottom electrode plates of the capacitor, ultimately resulting in changes in capacitance. Therefore, the value of the pressure can be calculated through the measurement of the change in capacitance.

Generally, in the preparation of graphene-based capacitive pressure sensors, there are three factors that affect the performance of capacitive pressure sensors, including membrane area, membrane thickness, and gap between top and bottom electrode plate of the capacitor. Therefore, large-area and high-quality suspended graphene membranes are critical for improving the sensitivity of pressure sensors. In 2016, Chen et al. proposed a novel method to fabricate large-area suspended graphene membranes for capacitive pressure sensors[181]. Their solvent substitution technique was used to fabricate graphene membranes up to 1.5 mm in diameter. The substitution process is illustrated in Fig. 9a, in which the PMMA/graphene/substrate is inverted and a bridging channel is formed so that acetone flowed through the channel to remove the PMMA layer. And in order to prevent rupture of the graphene membranes, the solvent was replaced by a solvent with a low surface tension after 10 min ($C_4F_9OH_3$). In addition to removing PMMA using the method above, thermal annealing was also used to remove the polymer layer, which gave twice the yield of the solvent substitution method and resulted in cleaner graphene samples. The sensitivity of the prepared graphene capacitive pressure sensor was 15.15 aF·Pa$^{-1}$, which is 770% higher than that of the conventional silicon-based pressure sensor, as shown in Fig. 9b.

In the fabrication of capacitive pressure sensors, a large parasitic capacitance is often generated, which can seriously affect the detection of the device capacitance. Thus, in order to improve the signal-to-noise ratio of pressure sensors, Dejan Davidovikj et al. proposed a novel method for static capacitive pressure sensing using graphene drums[182]. Their 3D schematic is illustrated in Fig. 9c. For one thing, the study used Au/Pd with lower surface roughness as the top electrode, which can provide better adhesion to the graphene membrane. And for another, a quartz substrate is used to reduce the parasitic capacitance, and due to the quartz substrate's excellent insulating properties, it can effectively isolate the capacitive coupling inside the device and between the devices. Therefore, the capacitance between the suspended graphene membrane and the bottom electrodes can be better detected rather than covered by the parasitic capacitance. Additionally, in order to improve the sensitivity of the pressure sensor, the distance between the top and bottom electrodes was reduced to 110 nm, and thus, capacitance changes as low as 50 aF could be detected.

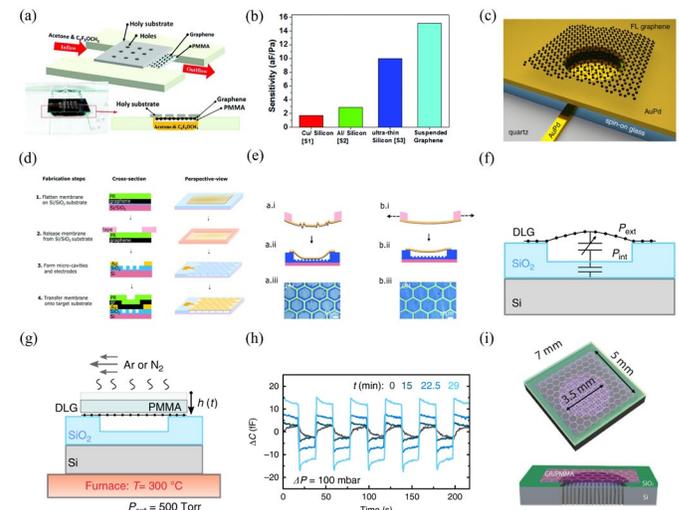

Fig. 9. Graphene-based capacitive gas pressure sensors. (a) Schematic illustration of the procedures involved in the solvent replacement method to remove the PMMA. (b) Comparison of the sensitivity of suspended graphene-based pressure sensor with other conventional pressure sensors. (c) Three-dimensional schematic of the graphene-based capacitive pressure sensor. (d) Schematic diagrams of the fabrication process of capacitive pressure sensor based on graphene-polymer. (e) Comparison of normal graphene membrane transfer and strained graphene membrane transfer. (f) Schematic cross-section of a graphene-based capacitive pressure sensor with graphene membrane deformed. (g) Schematic of the thermal annealing applied to reduce the thickness h(t) of PMMA. (h) Changes of capacitance under time-dependent pressure for samples with different annealing times. (i) The schematic of the cross-section of the graphene-based capacitive microphone. Adapted from [181], [182], [183], [165] and [166] under a Creative Commons license.

For capacitive pressure sensors, besides fulfilling the requirements for measuring slight pressure variations, it is also necessary to measure high ranges of pressure. Furthermore, during the formation of suspended graphene membrane, problems such as rupture of the graphene membrane can occur

8                                                                                                                                IEEE SENSORS JOURNAL, VOL. XX, NO. XX, MONTH X, XXXXdue to capillary or van der Waals forces in the graphene. Therefore, in 2017, Christian Berger et al. combined graphene with polymers to prepare a capacitive pressure sensor based on a heterostructure membrane with a high modulus of elasticity[183]. The preparation process is shown in Fig. 9d, and this heterostructure membrane is able to withstand large pressure variations while maintaining high sensitivity, thus increasing the measurement range of the sensor. Furthermore, the wrinkled graphene-polymer structure easily adheres to the bottom of the cavity during the transfer of graphene, so by applying a pre-strain to the graphene membrane during the transfer process, the membrane is flattened without adhering to the cavity and the local stress concentration of the membrane under pressure can be effectively reduced to enable the generated strain to be more uniform, thus improving the stability and accuracy of the pressure sensor output signal. The schematics of the samples without applied strain and with applied strain are shown in Fig. 9e. Likewise, using a suspended membrane array instead of single graphene drum to improve the sensitivity of capacitive sensors, Makars Šiškins et al. developed a capacitive pressure sensor based on double-layer graphene membrane array[165]. Its sensing cross-section is visualized in Fig. 9f. Furthermore, since the bending stiffness of the polymer introduced during the transfer of graphene reduces the deflection of the membrane, this study was carried out by removing the PMMA polymer in a dry gas (Ar or $N_2$) and in a high temperature environment (300 °C) to improve the performance of the pressure sensor. As shown in Fig. 9g, the thickness of the PMMA was gradually reduced with the increase of the thermal annealing time. The responsivity of the pressure sensor was increased by nearly an order of magnitude after 29 minutes of thermal annealing, indicating that removal of the polymer support layer by thermal treatment can further improve the sensitivity of the pressure sensor, as demonstrated in Fig. 9h. In addition to these improvements, this study developed a handheld, low-cost, battery-powered capacitive readout circuit capable of detecting pressure changes in static graphene drum structures. These innovations demonstrate significant advances in the design and optimization of graphene-based pressure sensors, particularly in improving sensitivity and developing portable, low-cost pressure sensors.

In fact, a sound sensor (e.g., a microphone) is essentially a pressure sensor. When sound waves reach the diaphragm of the microphone, the diaphragm will vibrate along with it, and this vibration changes the capacitance between the diaphragm and the fixed pole plate. Thus, this sound pressure value can be detected by detecting the change in capacitance. In 2021, Xu et al. developed an electrostatic microphone based on graphene/PMMA[166], whose cross-sectional schematic is shown in Fig. 9i. This is the first new example of successfully preparing large-area suspended graphene/PMMA membrane by releasing the silicon dioxide sacrificial layer. For acoustic pressure sensors, the diameter of graphene-based membranes is required to be in the order of several millimeters (3.5-7 mm), which is quite challenging for the transfer of such ultra-large-area graphene, and this study provides a method to transfer large-area graphene/PMMA membrane onto cavities.

### C. Graphene-Based Resonant Pressure Sensors

Resonant pressure sensors have the advantages of high stability, high repeatability, high accuracy, and high resolution. At present, less research has been conducted on resonant graphene pressure sensors. On the one hand, because the design and fabrication of resonant pressure sensors are usually more complicated than piezoresistive and capacitive sensors. On the other hand, resonant pressure sensors are significantly sensitive to environmental variations (e.g., temperature, humidity) and mechanical noises, which may affect their accuracy and stability.

In 2016, Robin J. Dolleman et al. proposed a graphene-based squeeze-film pressure sensor[167]. The cross-sectional view of this pressure sensor is shown in Fig. 10a. Different from capacitive pressure sensors, this sensor does not generate a pressure difference between the two sides of the graphene membrane. Instead, it increases the pressure inside the channel by compressing the gas inside the channel, which cannot escape from the channel in time due to its viscous force, thereby leading to the deformation of the graphene and the change of its resonance frequency. The critical innovation of this resonant pressure sensor is the use of few-layer graphene film. Due to the light weight and high flexibility of graphene, a large resonant frequency change (4 MHz) between 8 and 1000 mbar and the sensitivity of 9000 Hz/mbar were achieved, which is much higher than that of conventional MEMS pressure sensors (200 Hz/mbar). Foremost, the graphene-based squeeze-film pressure sensor does not require impermeable cavities, which can further reduce the size of the pressure sensors and at the same time is significant for the miniaturization of the sensors and the improvement of the sensitivity.

In 2017, S. Vollebregt proposed a squeeze-film pressure sensor based on suspended graphene film[184], the preparation process of this pressure sensor is shown in Fig. 10b. The size of the gap under the graphene beams can be accurately controlled by controlling the thickness of the catalyst Mo layer, which in turn affect the resonance frequency of the graphene beams. The experimental results show that the resonance frequency response of the sensor with a 100 nm gap is 310 Hz/mbar at low pressures (10-100 mbar) and 165 Hz/mbar at high pressures. In contrast, the device with a 200 nm gap has a lower resonance frequency response in these pressure ranges, as depicted in Fig. 10c. Due to the high Young's modulus of graphene, combined with the squeeze-film effect, the sensor can produce resonant frequency changes at smaller pressures, enabling high-precision measurement of gas pressure.

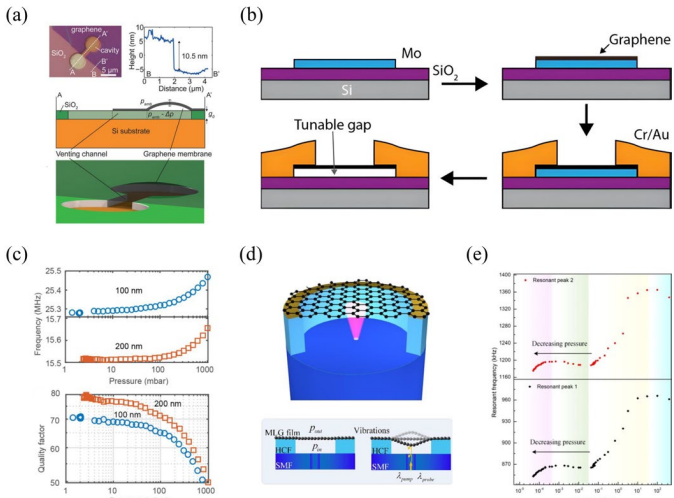

Fig. 10. Graphene-based resonant gas pressure sensors. (a) AFM image and schematic diagram of a graphene-based squeeze-film pressure sensor. (b) Fabrication process of the suspended graphene beams. (c) Resonant frequency and Quality factor versus pressure for the two different gap size devices. (d) Schematic diagrams of the clamped circular Au/graphene film-based pressure sensor. (e) Resonant frequency versus pressure for the fundamental and second resonant modes of a circular graphene membrane. Adapted from [167], [184] and [168] under a Creative Commons license.

Since most of the previous resonant pressure sensors are based on electrical readout systems, this invariably causes problems such as mechanical displacement noise, thermal noise, and electromagnetic interference, which in turn can interfere with the signal of the pressure sensor and significantly reduce the accuracy of the measurement. Therefore, in 2022, Chen et al. proposed a clamped circular Au/graphene film-based pressure sensor[168]. As can be seen from Fig. 10d, the space between the circular Au/graphene film and the fiber end facet forms a sealed cavity, and pressure variations affect the gas pressure inside this cavity, which in turn affects the film's tension and changes the film's resonance frequency. The pressure value can be detected by reading out the resonance frequency. The experimental data in Fig. 10e shows the performance of the sensor at different pressures, revealing the linear relationship between resonant frequency and pressure. This indicates that accurate measurement of pressure can be achieved by monitoring the change of the resonant frequency.

In addition, a significant advantage of this sensor over conventional mechanical resonant pressure sensors is that it detects pressure changes through resonant sensing with optical readout, which avoids mechanical displacement noise and parasitic capacitance problems present in electrical readout. And by monitoring higher order resonant modes, this pressure sensor can provide higher sensitivity, which provides a viable way to improve the performance of the pressure sensor.

### D. Other types of Suspended Graphene-Based Pressure Sensors

In addition to the piezoresistive, capacitive and resonant sensing mechanisms mentioned above, there are some other types of graphene-based pressure sensors, such as Pirani type. In 2018, Joost Romijn et al. proposed the world's first graphene-based Pirani pressure sensor[185]. As shown in Fig. 11a, the working mechanism of the Pirani pressure sensor is based on the heat conduction effect of gas molecules. When gas molecules come into contact with a graphene beam that generates Joule heat, they take away part of the heat of the graphene. And as the number of gas molecules increases (i.e., the gas pressure increases), this leads to a decrease in the heat and temperature of the graphene, and thus a change in the electrical resistance. By measuring the change of the electrical resistance of graphene, the change of the gas pressure can be measured. By growing graphene on Mo, problems such as membrane damage during the graphene transfer process are directly avoided. Meanwhile, the size of the cavity gap is regulated by controlling the thickness of $SiO_2$, thereby realizing the adjustment of the measured pressure range. As can be seen from the SEM image in Fig. 11b, this pressure sensor does not require a sealed cavity[185], which reduces the difficulty of process manufacturing. And the device can be better adapted to complex environmental conditions, such as high temperature, high humidity and other environments. In addition, due to the low thickness and high thermal conductivity of graphene, the sensor achieves a low power consumption of 0.9 mW and a resistance change of -2.75% compared to conventional Pirani pressure sensors[185].

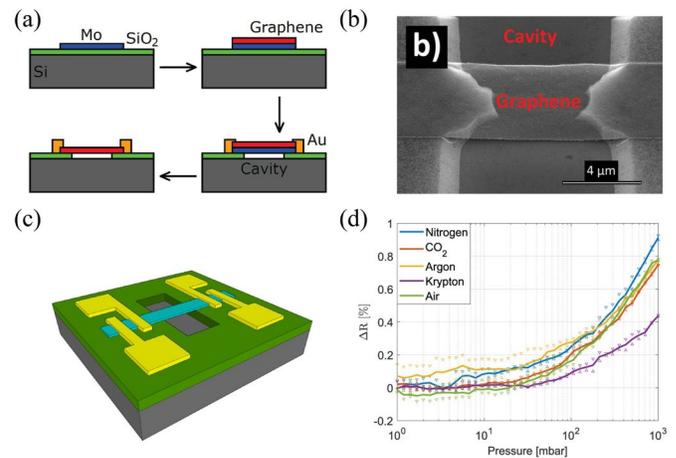

Fig. 11. Other types of graphene-based gas pressure sensors. (a) Fabrication process of the graphene-based Pirani pressure sensor. (b) SEM image of the suspended graphene beam over the etched cavity. (c) The schematic of 3D structure of the graphene-based Pirani pressure sensor. (d) The resistance change versus pressure of the graphene-based Pirani pressure under the conditions of different gases. Adapted from [185] and [186] under a Creative Commons license.

In 2021, Joost Romijn et al. further studied the graphene-based Pirani pressure sensor[186], whose three-dimensional structure is shown in Fig. 11c. And the response characteristics of graphene in the Pirani sensor to different gas molecules was studied. It can be seen from the Fig. 11d that there are obvious differences in the resistance changes caused by different gas environments, which indicates that the sensor has specific responses to different gas molecules. This is of great significance for the design optimization of the sensor.

In addition to graphene, other 2D materials also show potential applications in pressure sensing due to their unique physicochemical properties, such as transition metal sulfides ($MoS_2$, $MoSe_2$), MXene, and so on. In order to obtain a more comprehensive understanding of the current research status of pressure sensors based on 2D materials, TABLE IV summarizes the performance of pressure sensors based on other 2D materials in recent years, which will provide a broader vision and more diversified choices for the future research of pressure sensors.

TABLE IV
PERFORMANCE COMPARISON OF PRESSURE SENSORS BASED ON OTHER 2D MATERIALS.

| Materials | Sensitivity | Pressure range | Response time | Stability (cycles) | Year | Ref. |
|---|---|---|---|---|---|---|
| $PtSe_2$ | $1.05 \times 10^{-5}$ $kPa^{-1}$ | - | - | - | 2018 | [187] |
| $MoS_2$ | 89.75 $kPa^{-1}$ | 0 – 722.2 kPa | 3 ms | 5000 | 2019 | [188] |
| $MoS_2$ | 1036.04 $kPa^{-1}$ | 1 – 23 kPa | < 50 ms | 10000 | 2019 | [189] |
| $MoS_2$ | 2.21 $kPa^{-1}$ | 100 – 200 kPa | 200 ms | 1600 | 2021 | [190] |
| $MoSe_2$ | 18.42 $kPa^{-1}$ | 0.001 – 100 kPa | 110 s | 200 | 2021 | [191] |
| $MoSe_2$ | 0.3 $kPa^{-1}$ | 0.8 – 21 kPa | ~600 ms | 500 | 2021 | [192] |
| $MoSe_2/Ti_3C_2T_x$ | 14.7 $kPa^{-1}$ | 1.477 – 3.185 kPa | 385 ms | ~2500 | 2023 | [193] |
| MXene/$WS_2$ | 45.81 $kPa^{-1}$ | 0 – 410 kPa | 18 ms | 5000 | 2023 | [194] |
| $Ti_3C_2T_x$ | 6.31 $kPa^{-1}$ | 15 – 150 kPa | 300 ms | 2000 | 2022 | [195] |
| $Ti_3C_2T_x$ | 28.43 $kPa^{-1}$ | 0 – 8 kPa | 98.3 ms | 5000 | 2022 | [196] |
| MXene/PVA | 2320.9 $kPa^{-1}$(65.3 Pa-6.53 kPa) 1313.6 $kPa^{-1}$(6.53-65.3 kPa) 549.2 $kPa^{-1}$(65.3-294 kPa) | 65.3 Pa – 294 kPa | 70 ms | 10000 | 2022 | [197] |
| MXene/ANF | 120.64 $kPa^{-1}$(0 Pa-60 kPa) 521.69 $kPa^{-1}$(60 kPa-200 kPa) | 0 Pa – 20 kPa | 17 ms | 10000 | 2023 | [198] |
| MXene/BC/PDMS | 528.87 $kPa^{-1}$ | 0 – 10 kPa | 45 ms | 8000 | 2023 | [199] |
| MXene/AgNW | 770.86 $kPa^{-1}$ | 0 – 100 kPa | 70 ms | 5000 | 2023 | [200] |
| MXene/polyurethane | 150.65 $kPa^{-1}$ | 0.21 – 19.82 kPa | 75.5 ms | 10000 | 2024 | [201] |
| MXene nanosheets | 3.94 $kPa^{-1}$ | 0 – 117.5 kPa | 70 ms | 7500 | 2024 | [202] |

## V. CONCLUSIONS AND OUTLOOK

In summary, graphene and its derivatives have shown tremendous potential in the field of pressure sensors, including significant advances in the fields of flexible pressure sensors as well as gas pressure sensors. Firstly, the paper discusses in detail the excellent properties of graphene and its derivatives, including its common derivatives of graphene oxide and reduced graphene oxide. And at the same time, it discusses how these excellent properties can improve the performance of pressure sensors and facilitate the development of future technologies. Then the progress of graphene-based flexible pressure sensors and suspended graphene gas pressure sensors in recent years are provided in detail from the perspective of different sensing mechanisms, such as piezoresistive, capacitive, resonant, and piezoelectric, and their respective sensing performances are compared.

From the discussion of flexible pressure sensors, it can be concluded that the performance of flexible pressure sensors is currently affected by two main factors: the selection of the sensing material and the design of the sensing structure. Although many current flexible pressure sensors have demonstrated outstanding properties, they face many challenges in terms of their various performances, including sensitivity, detection pressure range, stability, and durability. For one thing, for flexible pressure sensors, achieving a better trade-off between sensitivity and detection range is still a current research challenge. And for another, there are still some deficiencies in the preparation process for suspended graphene pressure sensors. For example, the improvement of the graphene transfer method to achieve the integrity and cleanliness of the suspended graphene membrane and to reduce as much as possible the problems such as rupture and wrinkles that exist in the transfer is highly required, which is crucial for improving the performance of the pressure sensor. Moreover, the selection of the protective layer of the sensing material is also a critical factor affecting the sensitivity of the pressure sensor. Common protective layers of graphene include polymer (polydimethylsiloxane, PI; polymethylmethacrylate, PMMA), h-BN, $Si_3N_4$ nano-film, and other protective layers that not only prevent the graphene's oxidation and contamination but also ensure that the graphene still has good performance at high temperatures.

To be honest, graphene-based pressure sensors have the advantages and disadvantages. Some advantages of graphene-based pressure sensors are listed here:

(1) Graphene's high conductivity and thin layer thickness make the pressure sensor extremely sensitive to pressure, thus exhibiting high sensitivity, wide measurement range, and fast response time.
(2) The high strength and flexibility of graphene allows the pressure sensor to withstand large deformations without damage, resulting in high durability.
(3) The excellent performance of graphene-based pressure sensors at high temperatures makes them more suitable for operation in extreme environments.
(4) Compared to conventional silicon-based pressure sensors,

graphene-based pressure sensors can achieve nanoscale miniaturized dimensions while maintaining excellent performance, thus enabling device portability and high integration.

However, graphene-based pressure sensors also have some disadvantages due to some defects and deficiencies of graphene, such as:

(1) Low product yields due to the complexity and control difficulty of the preparation process, making it difficult to achieve large-scale production.
(2) Due to the extreme sensitivity of graphene to the environment, the graphene-based pressure sensors are easily affected by the environment (e.g., temperature, humidity, etc.) and need to be encapsulated.

With the continuous progress of science and technology and the cross-fusion of multiple disciplines, graphene-based pressure sensors will make important breakthroughs in various aspects. The following are a few outlooks for its future development trend:

(1) The future development of graphene-based pressure sensors will pay more attention to the optimization of materials and integration of functions. For example, by doping or compounding other functional nanomaterials, the mechanical properties, electrical conductivity and sensitivity of graphene can be further improved. Meanwhile, pressure sensing is integrated with other sensing functions (e.g., temperature, humidity, and gas sensing) to achieve the development of multifunctional sensors.
(2) Graphene's excellent flexibility and electrical conductivity make it promising for applications in flexible electronics and wearable devices. Graphene pressure sensors will develop in the direction of miniaturization and intelligence, and will be applied more to flexible displays, medical health monitoring, and flexible robots, etc., so that they can provide more portable and intelligent services for human life.
(3) In addition, pressure sensors will develop in the direction of energy self-sufficiency and intelligence. By combining with energy harvesting technologies (e.g., friction nanogenerators, thermoelectric converters), self-supply of pressure sensors will be realized to enhance their ability to work in unpowered environments.
(4) Pressure sensors can be combined with intelligent algorithms, including machine learning and deep learning algorithms, to enable advanced processing and analysis of collected data to provide more accurate monitoring results and predictions.

For future technological advances, the research on graphene pressure sensors should focus on solving the problems of the current manufacturing processing, such as the optimization of graphene transfer methods. And the reliability of graphene pressure sensors in extreme operating environments (e.g., high temperature and high pressure as well as environments with complex and harsh conditions) should be focused on. Therefore, how to select the appropriate protective materials to satisfy the requirements of the application environment to large extent is one of the critical research directions for graphene-based pressure sensors in the future. In addition, pressure sensors are currently developing in the direction of miniaturization and are committed to achieving a high degree of integration with IoT devices, wearable devices, and AI machines. Meanwhile, researchers should prioritize the scalability and cost-effectiveness of graphene materials and engage in more interdisciplinary collaborations that bring together expertise in materials science, mechanical engineering, electrical engineering, and computer science to achieve higher-performance pressure sensors. We expect this review will help researchers understand the development of graphene-based pressure sensors and promote the further development of graphene-based pressure sensors for future applications.

## Acknowledgment

Zhe Zhang, Quan Liu, Hongliang Ma and Ningfeng Ke are with the Advanced Research Institute for Multidisciplinary Science, Beijing Institute of Technology, Beijing 100081, China, also with the School of Integrated Circuits and Electronics, Beijing Institute of Technology, Beijing 100081, China.

Jie Ding is with the School of Integrated Circuits and Electronics, Beijing Institute of Technology, Beijing 100081, China. (e-mail: jie.ding@bit.edu.cn).

Wendong Zhang is with the State Key Laboratory of Dynamic Measurement Technology, North University of China, Taiyuan 030051, China, also with the National Key Laboratory for Electronic Measurement Technology, School of Instrument and Electronics, North University of China, Taiyuan 030051, China. (e-mail: wdzhang@nuc.edu.cn).

Xuge Fan is with the Advanced Research Institute for Multidisciplinary Science, Beijing Institute of Technology, Beijing 100081, China, also with the School of Integrated Circuits and Electronics, Beijing Institute of Technology, Beijing 100081, China, also with School of Physics, Beijing Institute of Technology, Beijing 100081, China. (e-mail: xgfan@bit.edu.cn).

(Corresponding authors: Xuge Fan, Jie Ding and Wendong Zhang)

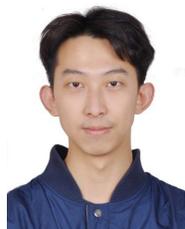

**Zhe Zhang** received his bachelor's degree from Taiyuan University of Technology in 2021. He is currently pursuing his master's degree in the Department of Electronic Information at Beijing Institute of Technology under the supervision of Prof. Xuge Fan.

His current research interests include micro-nano electromechanical sensors based on 2D materials such as graphene.

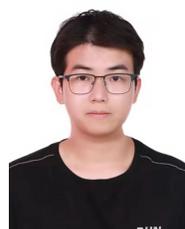

**Quan Liu** received his bachelor's degree from Taiyuan University of Technology in 2019. He is currently pursuing his master's degree at Beijing Institute of Technology under the supervision of Prof. Xuge Fan.

His current research interests include two-dimensional materials such as graphene and their sensing applications.

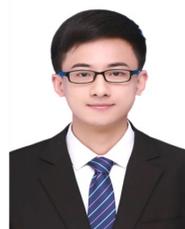

**Hongliang Ma** received his master's degree from Qilu University of Technology in 2020. He is currently a Ph. D. candidate at the School of Integrated Circuits and Electronics, Beijing Institute of Technology at Beijing, China, under the guidance of Prof. Xuge Fan.

His current research interests include the application of 2-dimensional material electronics in humidity and gas sensing.

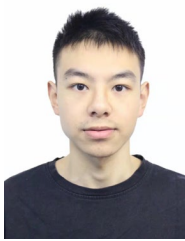

**Ningfeng Ke** received his Bachelor's Degree in Electronic Information Engineering from Beijing Institute of Technology in 2022. He is currently pursuing a Master's degree in Electronic Information Engineering at the advanced research institute of multidisciplinary sciences, Beijing Institute of Technology, China.

He is mainly engaged in research in the fields of MEMS/NEMS, 2D materials such as graphene, humidity sensors.

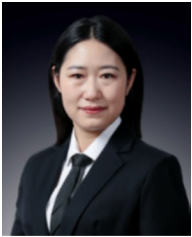

**Jie Ding** is an associate professor of Beijing Institute of Technology. She received her Ph. D. degree from University of Glasgow in 2015. She is currently an associate professor in School of Integrated Circuits and Electronics at Beijing Institute of Technology.

Her major research interests are advanced and novel semiconductor device and its modeling and circuit simulations.

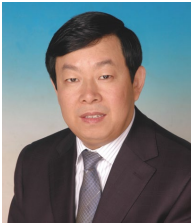

**Wendong Zhang** is a professor of North University of China. He received the Ph.D. degrees in 1995 from Beijing University of technology. He is currently a professor in School of Instrument and Electronics at North University of China, and the director of the State Key Laboratory of Dynamic Measurement Technology.

His major research interests are dynamic measurement technology, intelligent instruments and MEMS/NEMS.

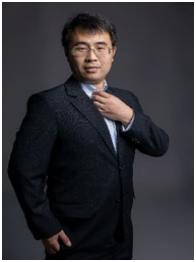

**Xuge Fan** is a professor of Beijing Institute of Technology. He received his Ph. D. degree from KTH Royal Institute of Technology in 2018.

His current research interests include graphene-based 2D materials, MEMS/NEMS and sensors.